\documentclass[aps,preprint]{revtex4}%
\usepackage{amsfonts}
\usepackage{amsmath}
\usepackage{amssymb}
\usepackage{graphicx}%
\begin{document}
\title{{\LARGE ENTROPY OF THE UNIVERSE}\\(to be published by Internation Journal of Theoretical Physics }

DOI 10.1007/s10773-009-9966-4)\author{Marcelo Samuel Berman$^{1}$}
\affiliation{$^{1}$Instituto Albert Einstein / Latinamerica - Av. Candido Hartmann, 575 -
\ \# 17}
\affiliation{80730-440 - Curitiba - PR - Brazil email: msberman@institutoalberteinstein.org}
\date{(Original: \textit{circa }2002; Last version: 8$^{th}$ December, 2008)}

\begin{abstract}
After a discussion on several limiting cases where General Relativity turns
into less sophisticated theories, we find that in the correct thermodynamical
and cosmological weak field limit of Einstein's field equations the entropy of
the Universe is $R^{3/2}$ -- dependent, where \ $R$\ \ \ stands for the radius
of the causally related Universe. Thus, entropy grows in the Universe,
contrary to Standard Cosmology prediction.

\end{abstract}
\maketitle

\begin{center}
\bigskip

{\LARGE ENTROPY OF THE UNIVERSE}

\bigskip(to be published by Internation Journal of Theoretical Physics 

DOI 10.1007/s10773-009-9966-4)

Marcelo Samuel Berman
\end{center}

\bigskip

\bigskip{\LARGE \bigskip I. Introduction}

\bigskip There are many different ways in order to compare General Relativity
with Newtonian Theory. In the next Section, we shall introduce a cosmological
weak field limit. In such limit, we can not ignore the existence of cosmic
pressure and a cosmological "constant" term. The possibility of introducing
cosmic pressure was surligned by Peacock (1999). Earlier, Ray d'Inverno (1992)
had worked with a cosmological repulsive acceleration in Newtonian Cosmology.
I recall Peter Landsberg and Evans (1977) having done something similar.

\bigskip

Other limits are possible, like the Machian (Berman, 2008a). A linearized
theory of the electromagnetic type, was also devised as some kind of Sciama's
limit to the field equations (Sciama, 1953; Berman, 2008b).

\bigskip

Barrow (1988), has worked what we shall call the thermodynamical Newtonian
"limit". He established the two energy conservation equations, namely, the
Newtonian and the thermodynamical; the latter implies the following definition
of pressure,

\bigskip

$p=-\frac{dM}{dV}$ \ \ \ \ \ \ \ \ \ \ \ \ \ \ \ \ \ ,

\bigskip

where \ \ $M$\ \ and \ \ $V$\ \ \ represent the mass and volume of a given
system. He showed that both equations led to Robertson-Walker's field
equations of General Relativity, provided that the energy density term,
instead of being related to mass energy, should be related to all forms of
energy, like, for instance, radiation.

\bigskip As to the Machian limits, Berman (2008a; 2007) has shown that there
are Newtonian-Machian and General Relativistic Machian limits, which, in fact,
result in the same conditions for the relevant Physical quantities.

\bigskip

We must deny, that the strong but wrong impression in the air, surrounding
Newtonian Cosmology as a pressureless model, is, in fact, true (Barrow, 1988).

{\LARGE II. Cosmological Newtonian limit of field equations}

Standard Cosmology (Weinberg, 2008), introduces constant entropy. We shall
show in next Section, that the Universe, could bear a growing entropy, in the
correct limit of Einstein's field equations.

\bigskip

It is well known that Einstein's field equations, in the so-called Newtonian
limit, reduce to Poisson's equation,

\bigskip

$\triangledown^{2}\Phi=4\pi G\sigma$ \ \ \ \ \ \ \ \ \ \ \ \ \ \ , \ \ \ \ \ \ \ \ \ \ \ \ \ \ \ \ \ \ \ \ \ \ \ \ \ \ \ \ \ \ \ \ \ \ \ \ \ \ \ \ \ \ \ \ \ \ \ \ \ \ \ \ \ \ \ \ \ \ \ \ \ \ \ \ \ \ \ \ \ \ \ \ \ \ \ (1)

\bigskip

where \ $\Phi$\ \ , \ \ $G$\ \ and \ \ $\sigma$\ \ \ stand for potential,
gravitational constant and gravitational energy density. What probably never
was told, is that \ $\sigma$\ \ \ represents not only the effective energy
density, but also eventual pressure and cosmological "constant" terms. In
fact, when applied in large scale, a pressure term may appear in the system,
and in a cosmological scale, a \ lambda term is possible. In
Whitrow's\ \ paper (Whitrow, 1946; Whitrow and Randall, 1951 ), he equated the
inertial energy of the Universe ( $Mc^{2}$\ ) to the gravitational potential
energy ( $G\frac{M^{2}}{R}$\ ), finding the approximate relation \ $G\frac
{M}{R}\cong c^{2}$\ \ .

\bigskip

If we postulate \underline{\textit{sphericity}}\textit{ }(the Universe
resembles a "ball" of approximate spherical shape), \underline
{\textit{egocentrism}} \ (each observer sees the Universe from its center) and
\underline{\textit{democracy}} (each point in space is equivalent to any other
one -- all observers are equivalent), we may write, for each observer, the
following Newtonian potential,

\bigskip

$\Phi=-G\frac{M}{R}$\ \ \ \ \ \ \ \ \ \ . \ \ \ \ \ \ \ \ \ \ \ \ \ \ \ \ \ \ \ \ \ \ \ \ \ \ \ \ \ \ \ \ \ \ \ \ \ \ \ \ \ \ \ \ \ \ \ \ \ \ \ \ \ \ \ \ \ \ \ \ \ \ \ \ \ \ \ \ \ \ \ \ \ \ \ \ \ \ \ \ \ \ \ (2)

\bigskip

\bigskip In the needed interpretation, we shall see \ $R$\ \ as the radius of
the causally related Universe. \ 

Then, \ from (1) and (2), we find,

\bigskip

$\triangledown^{2}\Phi=\frac{\partial^{2}\Phi}{\partial R^{2}}=-2\pi G\frac
{M}{R^{3}}=4\pi G\left(  \rho+3\frac{p}{c^{2}}-\frac{\Lambda}{4\pi G}\right)
$ \ \ \ \ \ \ \ \ \ \ \ \ \ \ . \ \ \ \ \ \ \ \ \ \ \ \ \ \ \ \ \ \ \ \ \ \ \ \ \ \ \ \ \ \ \ \ (3)

\bigskip

The obvious solution remains,

\bigskip

$\rho=\rho_{0}R^{-2}$ \ \ \ \ \ \ \ \ \ \ \ \ \ \ \ \ \ \ \ ,

\bigskip

$p=p_{0}R^{-2}$ \ \ \ \ \ \ \ \ \ \ \ \ \ \ \ \ \ \ \ , \ \ \ \ \ \ \ \ \ \ \ \ \ \ \ \ \ \ \ \ \ \ \ \ \ \ \ \ \ \ \ \ \ \ \ \ \ \ \ \ \ \ \ \ \ \ \ \ \ \ \ \ \ \ \ \ \ \ \ \ \ \ \ \ \ \ \ \ \ \ \ \ \ \ (4)

\bigskip

$\Lambda=\Lambda_{0}R^{-2}$ \ \ \ \ \ \ \ \ \ \ \ \ \ \ \ \ \ \ .

\bigskip

In the above, \ \ $\rho_{0}$\ \ \ , \ \ $p_{0}$\ \ and \ \ \ $\Lambda_{0}%
$\ \ \ \ are constants.\ The resulting equation is, for baryonic matter,

\bigskip

$M=\gamma$ $R\equiv-2\left[  \rho_{0}+3\frac{p_{0}}{c^{2}}-\frac{\Lambda_{0}%
}{4\pi G}\right]  $ $R>0$\ \ \ \ \ \ \ \ \ \ \ \ \ \ . \ \ \ \ \ \ \ \ \ \ \ \ \ \ \ \ \ \ \ \ \ \ \ \ \ \ \ \ \ \ \ \ \ \ \ \ \ (5)

\bigskip

\bigskip We show that negative pressures are possible. As dark matter is
represented by a positive cosmological term, for an accelerating Universe, we
have \ $\Lambda_{0}>0$\ \ . On the other hand, the weak energy condition,
stated as the positivity of energy density, should also apply and, thus,
\ $\rho_{0}>0$\ \ . From (5), we find that \ $p_{0}<\frac{\Lambda_{0}c^{2}%
}{12\pi G}-\frac{1}{3}\rho_{0}$ $c^{2}$ \ \ .

\bigskip We can check that Whitrow's relation is retrieved in its generality.

\bigskip{\LARGE III. Speaking \ of Thermodynamics }

We saw above, that energy densities are \ \ $R^{-2}$\ -- dependent. \ If all
energy densities are as such, we can say that the radiation component
\ $\rho_{rad}$\ \ has the same dependence, to wit,

\bigskip

$\rho_{rad}=b$ $R^{-2}=a$ $T^{4}$\ \ \ \ \ \ \ . \ \ \ \ \ \ \ \ \ \ \ \ \ \ \ \ \ \ \ \ \ \ \ \ \ \ \ \ \ \ \ \ \ \ \ \ \ \ \ \ \ \ \ \ \ \ \ \ \ \ \ \ \ \ \ \ \ \ \ \ \ \ \ \ \ \ \ \ \ \ \ \ \ (6)

\bigskip

In the above, \ \ $a$\ \ \ and\ \ $b$ \ are constants while \ \ $T$%
\ \ \ stands for absolute temperature, and the right hand side represents
black-body radiation. We find the same relation that Kolb and Turner (1990)
have already mentioned in another context, namely,

\bigskip

$T^{2}R=$ \ constant \ \ \ \ \ . \ \ \ \ \ \ \ \ \ \ \ \ \ \ \ \ \ \ \ \ \ \ \ \ \ \ \ \ \ \ \ \ \ \ \ \ \ \ \ \ \ \ \ \ \ \ \ \ \ \ \ \ \ \ \ \ \ \ \ \ \ \ \ \ \ \ \ \ \ \ \ \ \ \ \ \ \ \ \ \ \ (7)

\bigskip

In this case, the entropy of such Universe is given by (Sears and Salinger, 1975),

\bigskip

$S=S_{0}R^{3/2}$ \ \ \ \ \ \ \ ( \ $S_{0}$\ = constant ) \ \ \ \ \ \ \ . \ \ \ \ \ \ \ \ \ \ \ \ \ \ \ \ \ \ \ \ \ \ \ \ \ \ \ \ \ \ \ \ \ \ \ \ \ \ \ \ \ \ \ \ \ \ \ \ \ \ \ \ \ \ \ \ \ (8)

\bigskip

This entropy is growing, because \ $\dot{R}>0$\ \ for expanding Universes.
Weinberg (1972), suggested that there would be dissipative processes in the
Universe, that could be represented by viscosity terms.

\bigskip

{\LARGE IV. Final considerations}

\bigskip

\bigskip Some comments on the above Sections are necessary.

\bigskip

\textbf{first) }As a by-product of our presentation, we can check that there
is no infinite singularity in the beginning of this Universe, i.e., according
to (5),

\bigskip

$\lim\limits_{R\longrightarrow0}$ $M\longrightarrow0$
\ \ \ \ \ \ \ \ \ \ \ \ \ . \ \ \ \ \ \ \ \ \ \ \ \ \ \ \ \ \ \ \ \ \ \ \ \ \ \ \ \ \ \ \ \ \ \ \ \ \ \ \ \ \ \ \ \ \ \ \ \ \ \ \ \ \ \ \ \ \ \ \ \ \ \ \ \ \ \ \ \ \ \ \ \ \ \ \ \ \ \ \ (9)

\bigskip

\textbf{second) }\bigskip We have also found that pressure and density obey a
perfect gas equation of state, for both are \ $R^{-2}$\ -- dependent.\ As
dark-energy is representable by the lambda-term, this has also \ the same
dependence, which is consistent with Modern Cosmology (Weinberg, 2008).\ \ 

\bigskip

\bigskip\textbf{third) }The thermodynamical conclusion of ours, is also
confirmed by prior work on the above subjects, which was done by Berman (2007;
2007a; 2007b; 2008).

\bigskip

\bigskip\bigskip{\Large Acknowledgements}

\bigskip

First of all, I need to stress the importance of an anonymous referee's
report, whose comments directed me to write this version of the paper. I thank
Nelson Suga, Marcelo F. Guimar\~{a}es, Antonio F. da F. Teixeira, and Mauro
Tonasse; I am also grateful for the encouragement by Albert, Paula, and Geni.
I offer this paper \textit{in memoriam of \ \ }M. M. Som

\bigskip

{\Large References}

\bigskip

\bigskip Barrow, J.D. (1988) - \textit{The Inflationary Universe, }R. Blin
Stoyle \& W.D. Hamilton (eds), Adam Hilger, Bristol. Paper presented at the
University of Sussex, date 7-9$^{th}$ September, 1987, on \textit{Interactions
and Structures in Nuclei.}\ \ 

Berman,M.S. (2007) - \textit{The Pioneer Anomaly and a Machian Universe,
}Astrophysics and Space Science, \textbf{312}, 275.

Berman,M.S. (2007a) - \textit{Introduction to General Relativity and the
Cosmological Constant Problem}, Nova Science, New York.

Berman,M.S. (2007b) - \textit{Introduction to General Relativistic and
Scalar-Tensor Cosmologies}, Nova Science, New York.

\bigskip Berman,M.S. (2008) - \textit{A Primer in Black-Holes, Mach's
Principle, and Gravitational Energy }, Nova Science, New York.

\bigskip Berman,M.S. (2008a) - \textit{General Relativistic Machian Universe,
}Astrophysics and Space Science, \textbf{318}, 273-277.

Berman,M.S. (2008b) - \textit{On the Machian Origin of Inertia, }Astrophysics
and Space Science, \textbf{318}, 269-272.

d'Inverno, R. (1992) - \textit{Introducing Einstein's Relativity, }Clarendon
Press, Oxford.

Kolb, E.W.; Turner, M.S. (1990) - \textit{The Early Universe}, Addison-Wesley, N.Y.

Landsberg, P. T.; Evans, D.A. (1977) - \textit{Mathematical Cosmology, }Oxford
Universe Press, Oxford.

Peacock, J.A. (1999) - \textit{Cosmological Physics, }Cambridge Universe
Press, Cambridge, page 25, formula (1.85).

Sciama, D.N. (1953) - MNRAS, \textbf{113,} 34.

Sears, F.W.; Salinger,G.L. (1975) - \textit{Thermodynamics, Kinetic theory,
and Statistical Thermodynamics, }Addison-Wesley, New York.

Weinberg, S. (1972) - \textit{Gravitation and Cosmology, }Wiley, New York.

Weinberg, S. (2008) - \textit{Cosmology,} Oxford University Press, Oxford.

Whitrow, G. (1946) - Nature, \textbf{158}, 165.

Whitrow, G.; Randall, D. (1951) - MNRAS, \textbf{111}, 455.

\end{document}